\documentclass[amssymb,twocolumn,10pt,a4paper]{extarticle}
\usepackage{amsmath}
\usepackage{amssymb}
\usepackage{physics}
\usepackage{mhchem}
\usepackage{geometry}
\usepackage{graphicx}
\usepackage{subcaption}
\usepackage{siunitx}
\usepackage[super,comma,sort&compress]{natbib}
\usepackage{hyperref}
\usepackage{cleveref}
\usepackage{xcolor}
\usepackage{authblk}

\begin{document}
\title{Electrically Reconfigurable Silicon Carbide Nanophotonic Cavities on Thin-Film Lithium Niobate}

\author{San Lam Ng*}
\date{*Email: \href{mailto:san-lam.ng@pi3.uni-stuttgart.de}{san-lam.ng@pi3.uni-stuttgart.de}}

\author{Georgii Grechko}
\author{Vladislav Bushmakin}
\author{Rainer Stöhr}
\author{Vadim Vorobyov}
\author{Roman Kolesov}
\author{Jörg Wrachtrup}
\affil{The 3rd Institute of Physics, University of Stuttgart, Germany}

\twocolumn[\begin{@twocolumnfalse}
	\maketitle
	\begin{abstract}
		Interfacing integrated photonics with solid-state spin defects holds great promise for future quantum networks, but the scaling of spin-photon architectures is hindered by frequency mismatches arising from fabrication-induced variations in photonic cavity resonances and the inhomogeneous optical transition frequencies of individual spins. These challenges call for a photonic platform with deterministic and wide-range tunability. Here, we demonstrate a hybrid nanophotonic platform based on direct bonding of silicon carbide photonic crystal nanocavity arrays onto thin-film lithium niobate on insulator, enabling deterministic electrical tuning of multiple SiC nanocavities into mutual spectral resonance. By exploiting the strong electro-optic response of lithium niobate, we achieve continuous and wide-range cavity tuning of 380 GHz ($\sim$1.1 nm), sufficient to compensate both cavity disorder and spin inhomogeneity. The nanocavities balance strong optical confinement with electrical tunability, exhibiting a theoretical Purcell factor of approximately 400. This hybrid platform enables electrically reconfigurable spin-photon interfaces for large-scale integrated quantum photonics.
		\newline
	\end{abstract}
\end{@twocolumnfalse}]

\section{Introduction}
	\label{sec:introduction}
	Solid-state spin defects constitute a leading platform for quantum networks, as they combine long-lived electron and nuclear spin coherence with optically addressable transitions suitable for efficient spin-photon interfaces. These interfaces underpin key functionalities including quantum communication, distributed entanglement generation, and quantum repeater architectures. Over the past two decades, a variety of solid-state spin systems have been developed as individual quantum nodes, notably nitrogen-vacancy and group-IV vacancy centers in diamond \cite{10.1557/mrs.2013.20}, silicon- and divacancy-related defects in silicon carbide (SiC) \cite{10.1103/PRXQuantum.1.020102,10.1088/2515-7647/ab77a2}, and rare-earth ions embedded in crystalline hosts \cite{10.1002/lpor.202300257} such as yttrium silicate. Despite substantial progress at the level of single devices, extending these systems to large-scale, interconnected networks remains a central challenge.

	Among the available host materials, SiC has emerged as a particularly compelling platform for scalable spin-photon architectures. In addition to supporting multiple optically addressable spin defects with long coherence times, SiC benefits from mature wafer-scale growth, compatibility with established semiconductor fabrication processes, and the ability to co-integrate photonic, electronic, and mechanical elements within a single material system \cite{10.1038/s41378-023-00496-1,10.1364/PRJ.567674}. Furthermore, SiC hosts spin defects with optical transitions spanning the visible to near-infrared spectral range \cite{10.1088/2515-7647/ab77a2}, enabling flexible integration with nanophotonic structures. These attributes make SiC uniquely suited for scalable quantum photonic systems requiring large arrays of reproducible, individually addressable spin-photon nodes.

	A key obstacle to such scalability, however, lies in the deterministic spectral alignment of spin optical transitions with nanophotonic cavity modes. Nanophotonic cavities rely on subwavelength feature sizes, rendering their resonance frequencies highly sensitive to fabrication imperfections that inevitably introduce device-to-device variations. In parallel, solid-state spin defects exhibit intrinsic inhomogeneity in their optical transition frequencies arising from local strain, charge environment fluctuations, and crystal-field variations. As a result, the spectral mismatch between cavity resonances and spin transitions typically exceeds the natural linewidths of both systems, severely limiting the yield of usable spin-cavity devices and precluding straightforward scaling to large arrays.
	\begin{figure*}[htb]
		\centering
		\includegraphics[width=\textwidth]{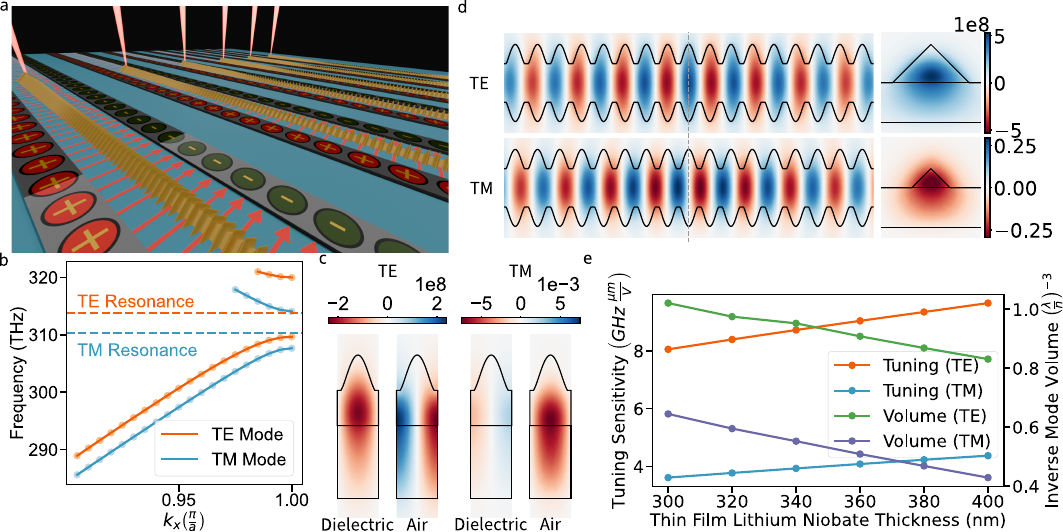}
		\caption{\textbf{a}. Schematic illustration of the cavity design. \textbf{b}. Simulation of the cavity bandgap, both resonance modes are placed at the middle of the bandgap. \textbf{c}. Simulation of the in-plane component of a unit cell. \textbf{d}. Simulation of the in-plane component of the resonance modes. The left figures show the top view, with dash line indicating the position of cavity centre. The right figures show the cross section view at the positions of field maxima. \textbf{e}. Simulated dependence of tuning sensitivity and mode volume on thin film LN thickness, showing that either the mode volume or the tuning sensitivity could be optimized depending on the need.}
		\begin{subfigure}{0.0\textwidth}
			\phantomcaption{}
			\label{fig:cavity_model}
		\end{subfigure}
		\begin{subfigure}{0.0\textwidth}
			\phantomcaption{}
			\label{fig:simulation_bandgap}
		\end{subfigure}
		\begin{subfigure}{0.0\textwidth}
			\phantomcaption{}
			\label{fig:simulation_field_unitcell}
		\end{subfigure}
		\begin{subfigure}{0.0\textwidth}
			\phantomcaption{}
			\label{fig:simulation_field_resonance}
		\end{subfigure}
		\begin{subfigure}{0.0\textwidth}
			\phantomcaption{}
			\label{fig:tfln_thickness_effect}
		\end{subfigure}
		\label{fig:summary}
	\end{figure*}

	Post-fabrication, in-situ tuning of cavity resonances is therefore essential. In SiC-based nanophotonic systems, cavity tuning has predominantly relied on gas adsorption techniques \citep{10.1063/1.2076435}, which provide a limited tuning range and suffer from slow response times, lack of reversibility, and unreliable independent control of individual cavities. Alternative tuning approaches explored across nanophotonics, including thermal tuning \citep{10.1109/LPT.2004.826781}, mechanical strain \cite{10.1063/1.1649803}, and electro-optic modulation \citep{10.1038/nphoton.2007.93,10.1038/s41467-023-37513-w,10.1364/OPTICA.453527} offer different trade-offs in speed, stability, and scalability. Among these, electrical tuning is particularly attractive for quantum network applications, as it enables fast, deterministic, and potentially independent control of large numbers of devices. However, the weak electro-optic response of SiC severely limits the achievable tuning range in purely SiC photonic devices. Hybrid approaches that integrate SiC with materials possessing strong electro-optic effects thus offer a promising route to enable efficient electrical tuning while preserving the favourable spin and fabrication properties of SiC.

	We propose and experimentally demonstrate a hybrid photonic architecture in which SiC photonic crystal cavity arrays are integrated onto thin-film lithium niobate (TFLN). In this configuration, the optical cavity mode extends from the SiC nanophotonic structure into the TFLN layer, forming a cavity jointly defined by the two materials. Owing to its large electro-optic coefficient, LN enables efficient electrical tuning, while SiC hosts the spin defects, allowing a spatial separation between tuning functionality and the defect spin environment. By balancing tuning efficiency with optical mode confinement, this approach extends electrical control to spin-hosting materials that lack intrinsic electro-optic response while preserving favourable spin properties.
	We realize this architecture using designed corrugated nanobeam cavities with a triangular cross-section transferred onto TFLN. The devices are fabricated by creating near-free-standing SiC cavities, followed by their release and bonding onto LNOI, allowing deterministic and large-scale array integration. Taking 4H-SiC as a representative host material, we achieve electrical tuning of the cavity resonances over 380 GHz ($\sim$1.1 nm). Furthermore, we demonstrate electrical alignment of multiple cavities to a common resonance frequency, providing a key step toward scalable and electrically reconfigurable integrated quantum photonic platforms.

	\begin{figure*}
		\centering
		\includegraphics[width=\textwidth]{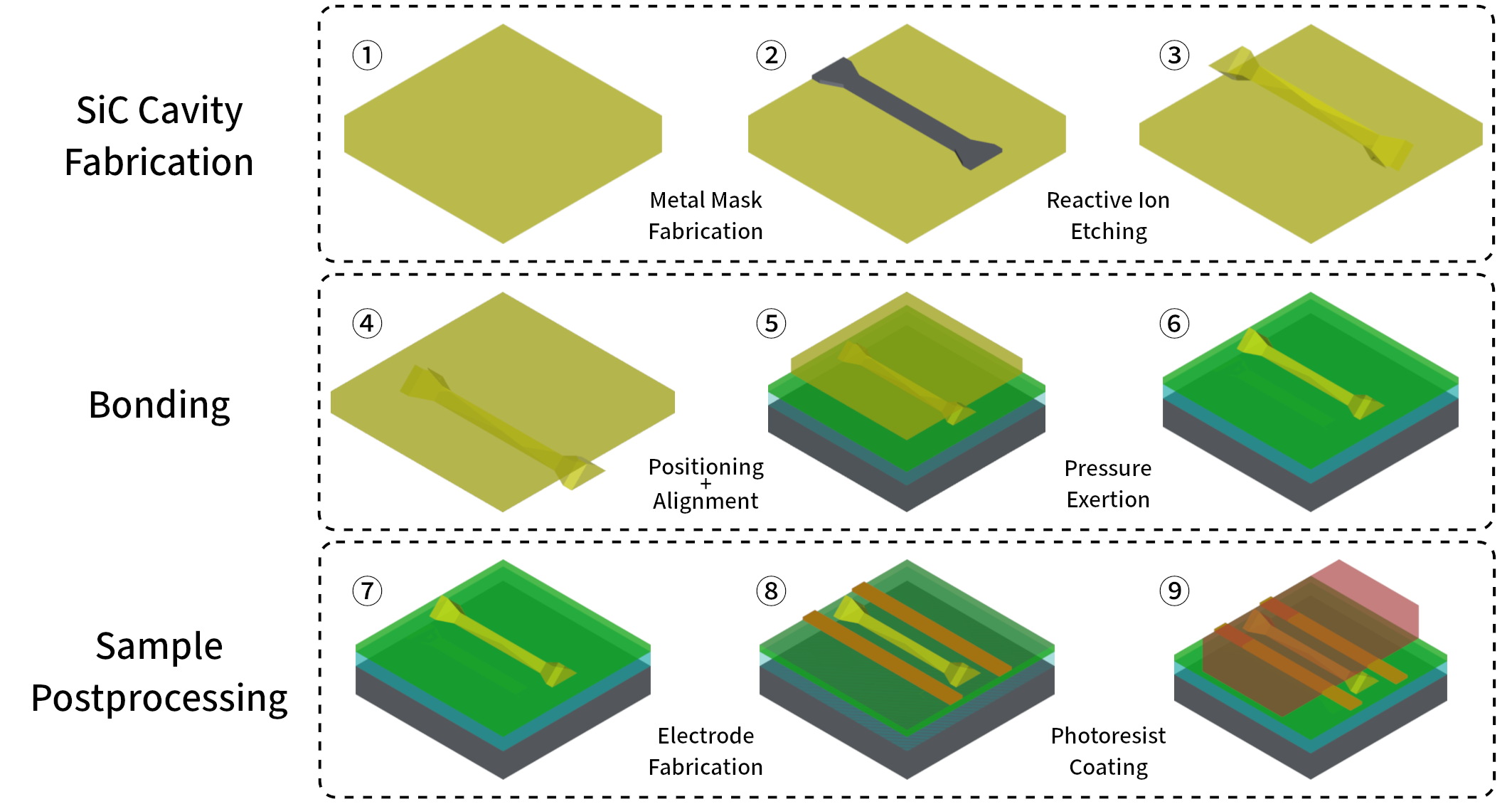}
		\caption{\textbf{Outline of the fabrication procedure.} The fabrication process is loosely divided into three parts. 1. Fabrication of triangular corrugated cavities from 4H-SiC by e-beam lithography followed by e-beam metal evaporation and reactive ion etching. 2. Bonding of nanobeams to LNOI by direct bonding. 3. Electrode fabrication and protective cladding of cavities.}
		\label{fig:nanofabrication_steps}
	\end{figure*}

\section{Cavity Design}
	\label{sec:design}
	The hybrid cavity comprises two portion, a triangular corrugated cavity beam made from 4H-SiC on top and a flat layer of TFLN underneath, as depicted in figure \labelcref{fig:cavity_model}. We designed this geometry as a balance between experimental simplicity (see also section \ref{sec:fabrication}) and cavity performance. With the reflection symmetry naturally preserved, the cavity resonance modes are separated into TE and TM modes, resulting in simpler theoretical and experimental treatment.

	The cavity beam portion serves to provide light confinement as in a regular cavity. We swept the unit cell period $a_0$ for the Bragg reflectors from $\SI{220}{\nm}$ to $\SI{230}{\nm}$ to target the infra-red. The inner and outer widths of the corrugation were chosen to be $w_l = \SI{385}{\nm}$ and $w_h = \SI{770}{\nm}$, ensuring sufficiently wide bandgaps for both modes, as shown in figure \ref{fig:simulation_bandgap}. We created the cavity using a Gaussian modulation of cell period, described by $\frac{\Delta a}{a_0} = - A e^{- \left(\frac{n}{\sigma}\right)^2}$, with $A = 6.5 \%$, $\sigma = 5$. The chosen function resulted in the resonance modes sharing the characteristics of the dielectric band as illustrated in figure \labelcref{fig:simulation_field_unitcell,fig:simulation_field_resonance}. The chosen geometry yields mode volumes of $1.3 \left(\frac{\lambda}{n}\right)^3$ and $2.4 \left(\frac{\lambda}{n}\right)^3$ for TE and TM mode respectively. For facilitation of light coupling, we included a segment of waveguide for each end, the waveguides are linearly tapered to minimize loss, and we further incorporated a short section of support to hoist and anchor the beams to the bulk.

	Underneath the beam cavity is the LN layer, which enables electrical tuning of the hybrid cavity. We chose to use TFLN over bulk LN as the \ce{SiO_2} layer provides index confinement to the hybrid cavity. The thin film thickness controls the mode location in the two materials, and hence the trade off between tuning sensitivity and mode volume. We opted for a thickness of $\SI{400}{\nm}$ for the current work, and further improvement to mode volume and hence Purcell factor could be expected by thinning the TFLN. For maximal tuning range, we align the z-axis of the LN with the electric field of both TE mode and the external field by the electrode.

\section{Fabrication}
	\label{sec:fabrication}
	As outlined in figure \ref{fig:nanofabrication_steps}, we first create the corrugated cavity portion of the hybrid cavity from bulk 4H-SiC (see supplementary information for details of the source materials). The 2D geometry at the base of the corrugated cavity was created by e-beam lithography using bilayer PMMA resist. Then, Nichrome was deposited as etch mask using e-beam evaporation. We then utilized an angle etching technique \citep{10.1063/1.91554,doi:10.1021/nl302541e} (see supplementary information for details) for reactive ion etching to create the suspended triangular cavity beam with only two thin supports at the ends.

\begin{figure*}
		\centering
		\includegraphics[width=\textwidth]{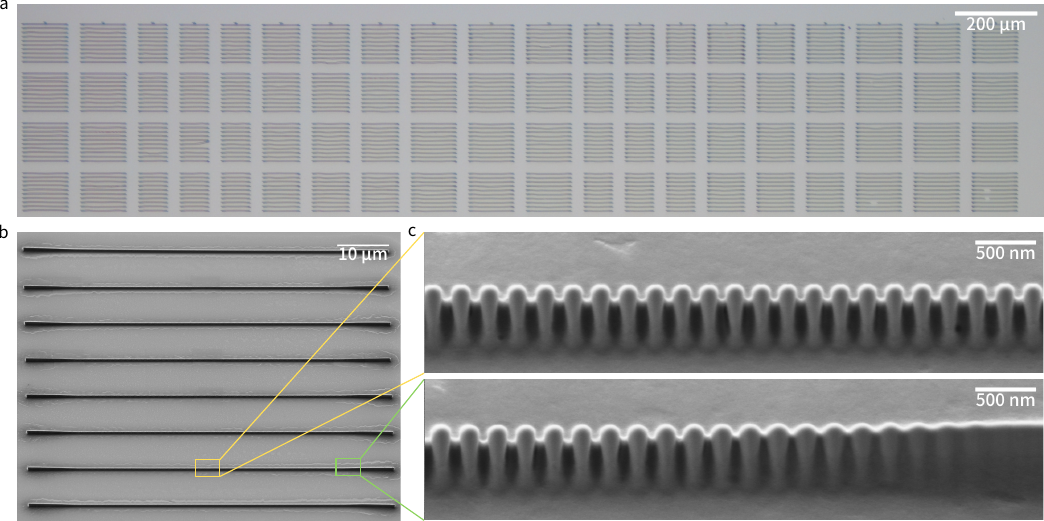}
		\begin{subfigure}{0.0\textwidth}
			\phantomcaption{}
			\label{fig:bonding_optical_goodyield}
		\end{subfigure}
		\begin{subfigure}{0.0\textwidth}
			\phantomcaption{}
			\label{fig:bonding_sem_array}
		\end{subfigure}
		\begin{subfigure}{0.0\textwidth}
			\phantomcaption{}
			\label{fig:bonding_sem_zoomin}
		\end{subfigure}
		\caption{\textbf{Images of bonded cavities.} \textbf{a}. Optical image of a large matrix of bonded cavities of a high yield sample, demonstrating the efficiency of the bonding method. \textbf{b}. SEM image of an array of bonded cavities. \textbf{c}. Defect and taper sections of one of the cavity.}
		\label{fig:bonding_results}
	\end{figure*}

	After creation of the beam cavities, using oxygen plasma, we cleaned and activated both 4H-SiC and LN surfaces. We then attach the 4H-SiC beam cavities to the TFLN surface by direct bonding to complete the hybrid cavities. In the bonding process, the two bulks chips were brought into contact using a micro-manipulator, pressure was then exerted using a manipulator arm to break the thin cavity beam supports and bond the 4H-SiC cavity beams to TFLN. The use of the ``break and bond'' method with the aid of micro-manipulator results in a good yield as shown in figure \labelcref{fig:bonding_optical_goodyield,fig:bonding_sem_array}, and the cavities retained good surface profile as shown in figure \labelcref{fig:bonding_sem_zoomin}. After bonding, the hybrid cavities were annealed at $325^{\circ}C$ for 90 minutes to enhance the bonding strength.

	Following bonding, we created the electrodes at about $\SI{3}{\mu m}$ from the 4H-SiC cavity beams using optical lithography followed by e-beam evaporation of gold. We completed the fabrication process by coating the body of the cavities with a layer of photoresist with low absorption in the infra-red range for protection against dielectric breakdown.

\section{Results}
	\label{sec:results}
	Cavities were characterized using a home-built bi-focal microscope (see supplementary information for details of the setup).
	\begin{figure*}[htb]
		\centering
		\includegraphics[width=\textwidth]{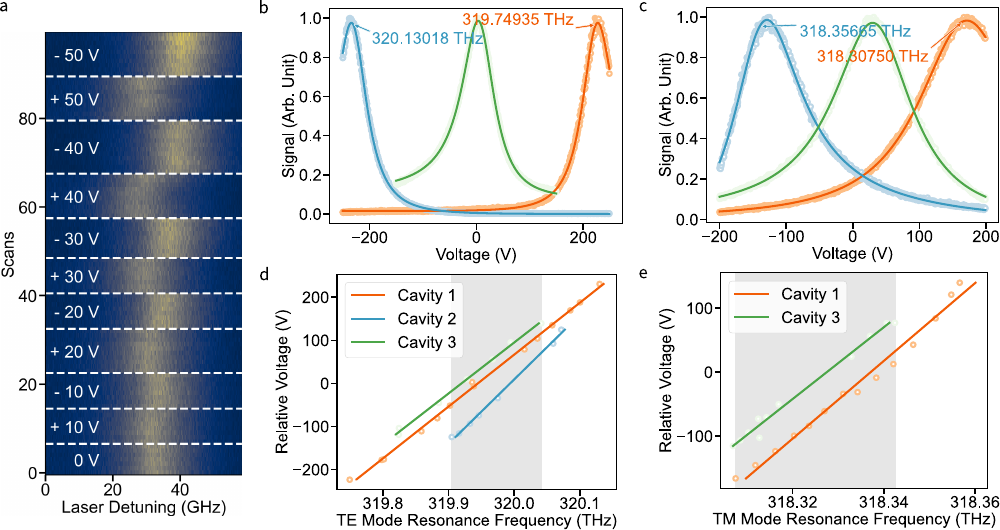}
		\begin{subfigure}{0.0\textwidth}
			\phantomcaption{}
			\label{fig:result_laser_sweep}
		\end{subfigure}
		\begin{subfigure}{0.0\textwidth}
			\phantomcaption{}
			\label{fig:result_te_max_tuning}
		\end{subfigure}
		\begin{subfigure}{0.0\textwidth}
			\phantomcaption{}
			\label{fig:result_tm_max_tuning}
		\end{subfigure}
		\begin{subfigure}{0.0\textwidth}
			\phantomcaption{}
			\label{fig:result_tuning_sensitivity_te}
		\end{subfigure}
			\begin{subfigure}{0.0\textwidth}
			\phantomcaption{}
			\label{fig:result_tuning_sensitivity_tm}
		\end{subfigure}
		\caption{\textbf{Results of electrical tuning tests.} \textbf{a}. Laser scattering intensity as a function of frequency detuning for the TM mode of the cavity. Voltages were applied to the cavity in steps, and shifts in the resonance frequency were observed, verifying the successful tuning of cavity resonance. \textbf{b,c}. Voltage sweep performed on TE and TM mode of the cavity, confirming a tuning range of $\SI{380.8}{\GHz}$ for TE and $\SI{49.1}{\GHz}$ for TM mode. \textbf{d,e}. Tuning sensitivity ($\SI{}{\GHz \per \V}$) of TE and TM modes of characterized cavities, the shaded region highlight the common spectral range among the tested cavities.}
		\label{fig:experimental_results}
	\end{figure*}
	We first qualitatively confirmed the electrical tuning of cavities, by applying an external voltage in steps while continuously sweeping the laser frequency around the TM mode resonance. Abrupt jumps in the resonance were observed that were caused by steps in applied voltage, as shown in figure \ref{fig:result_laser_sweep}, thereby confirming the electrical tuning of the cavities. 

	For quantitative characterization of the cavities, we fixed laser frequency at some values around the resonance and swept the applied voltage to bring the laser in resonance with the cavity. This allowed us to scan a much wider range than the mode-hop-free scan range of the ECDL. In order to assess the tuning capabilities of the hybrid cavity, up to $\SI{\pm 250}{\V}$ ($\SI{\pm 200}{\V}$) was applied to the TE (TM) mode of a cavity. We were able to directly confirm achievable tuning ranges of $\SI{380.8}{\GHz}$ and $\SI{49.1}{\GHz}$ for TE and TM mode respectively, as shown in figures \labelcref{fig:result_te_max_tuning,fig:result_tm_max_tuning}. The demonstrated tuning range is sufficient to address \ce{V_2} centres \cite{doi:10.1021/acs.nanolett.4c02162} and some other spin defects \cite{10.1063/5.0077112,10.1038/s41534-018-0097-8} across their inhomogeneous broadening.

	We then characterized a number of cavities of the same geometrical design using lower maximum voltages. As figures \labelcref{fig:result_tuning_sensitivity_te,fig:result_tuning_sensitivity_tm} show, we were able to tune the cavities into a common resonance, indicated by the shaded regions in figures, demonstrating that electrical tuning of the hybrid cavity could address the manufacturing variance inherent to nanofabrication. The ability to address both variation in spin and cavity resonances suggests that the hybrid cavity platform provides a viable avenue towards scalable creation of spin-photon interfaces.

	Across the characterized cavities, we measured tuning parameters of up to $\SI{850}{\MHz \per \V}$ ($\SI{178}{\MHz \per \V}$) and $\SI{783}{\MHz \per \V}$ ($\SI{171}{\MHz \per \V}$) on average for TE (TM) mode. Converted to electric field, the parameters were \emph{estimated} to be between $4.5 - \SI{5.7}{\GHz \um \per \V}$ and $1.1 - \SI{1.6}{\GHz \um \per \V}$, lower than $\SI{9.7}{\GHz \um \per \V}$ and $\SI{4.4}{\GHz \um \per V}$ expected from simulation. We speculate that accumulated surface charges might have partially shielded the cavities from the applied field, or that there exists lot to lot variation in electro-optics coefficient of the TFLN wafer, or that LN might have been poled, diminishing the effective electro-optics coefficient, due to the high electric field applied possibly exceeding coercive field strength of LN. Further work is needed to understand and possibly mitigate the discrepancy. 

	\begin{figure*}[htb]
		\centering
		\includegraphics[width=\textwidth]{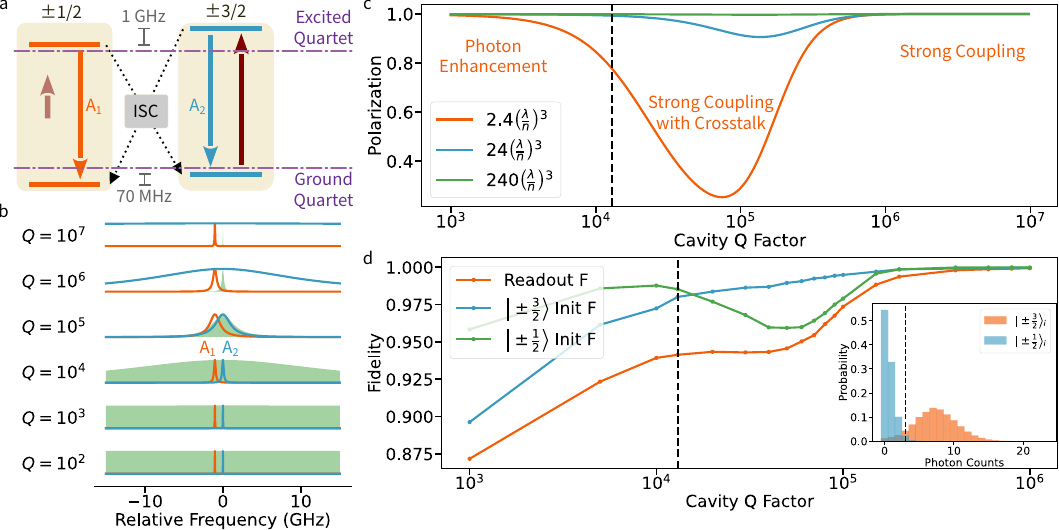}
		\begin{subfigure}{0.0\textwidth}
			\phantomcaption{}
			\label{fig:v2_energylevel}
		\end{subfigure}
		\begin{subfigure}{0.0\textwidth}
			\phantomcaption{}
			\label{fig:optical_broadening}
		\end{subfigure}
		\begin{subfigure}{0.0\textwidth}
			\phantomcaption{}
			\label{fig:optical_initialization_fidelity}
		\end{subfigure}
		\begin{subfigure}{0.0\textwidth}
			\phantomcaption{}
			\label{fig:singleshot_fidelity}
		\end{subfigure}
		\caption{\textbf{\emph{Estimated} cavity enhancement for \ce{V_2} centres under optimal placement.} \textbf{a}. Simplified energy levels structure of the spin degree of freedom for \ce{V_2} centre in the absence of external magnetic field \cite{10.1038/s41534-024-00861-6}. Solid arrows denote radiative transitions and dashed arrows denote non-radiative decay channel. \textbf{b}. Illustration of linewidth broadening of the two electron spin state selective transitions. Blue (orange) curve represents \ce{A_2} (\ce{A_1}) transition and the green shade represents the spectral width of the resonance mode. The cavity mode is in resonance with the \ce{A_2} transition. At both low and very high Q factor, one or both transitions remain narrow, allowing the on-resonance transition to be selectively excited. Around $Q \sim 10^{4} - 10^{5}$, both transitions are appreciably broadened. \textbf{c}. Achievable spin polarization ($\ket{\pm \frac{1}{2}}$ subspace) by resonant excitation of \ce{A_2} transition. The dashed line indicates the achieved Q factor. The achievable polarization drops than rises with increasing Q factor, due to broadening and excitation of \ce{A_1} transition. For comparison, the achievable polarization at 10 times (for micro Fabry-Perot cavity, ring resonator) and 100 times (ring resonator) of the cavity mode volume are depicted. \textbf{d}. Achievable readout and initialization fidelities by single-shot readout and measurement based initialization. The dashed line indicates the achieved Q factor. Inset shows an example histograms of the photon distribution for $Q = \SI{13e3}{}$, where the vertical line indicates the optimal readout discrimination threshold.}
		\label{fig:cavity_enhancement}
	\end{figure*}

	From the measured traces, the cavity linewidths and hence the Q factors were determined,  tabulated in table \ref{tab:estimated_cavity_q_factor}. We observed Q factor up to $\SI{6e3}{}$ for TE mode and up to $\SI{13e3}{}$ for TM mode, comparable with reported values for pure 4H-SiC cavities. This suggests that the hybrid integration approach had not significantly impacted the cavity performance, and the Q factor might be limited by the high doping concentration of the 4H-SiC sample \cite{10.1364/OE.27.013053,10.1364/OL.40.004138}.
	\begin{table}[htb]
		\centering
		\begin{tabular}{c|c|c}
			\hline
			Cavity & Mode & \emph{Estimated} Q Factor
			\\ \hline
			Cavity 1 & TE & $\sim 5 \times 10^{3}$
			\\ \hline
			Cavity 1 & TM & $\sim 13 \times 10^{3}$
			\\ \hline
			Cavity 2 & TE & $\sim 6 \times 10^{3}$
			\\ \hline
			Cavity 3 & TE & $\sim 6 \times 10^{3}$
			\\ \hline
			Cavity 3 & TM & $\sim 12 \times 10^{3}$
			\\ \hline
		\end{tabular}
		\caption{\emph{Estimated} Q factors for the various cavity modes tested.}
		\label{tab:estimated_cavity_q_factor}
	\end{table}

	In order to understand the optical characteristics of a spin embedded in the hybrid cavity and by extension, the prospect of realizing a spin-photon interface, we computed expected cavity enhancement parameters of a \ce{V_2} centre under optimal placement (details provided in supplementary information). We calculate the parameters of the TM mode assuming that the optical dipole of \ce{V_2} centre is aligned with the field direction of TM mode, the values are tabulated in table \ref{tab:cavity_enhancement}. Notably, we obtain a theoretical overall Purcell factor of up to $410$, which translates to an effective Purcell factor of $11.2$ ($21.0$) for the $\ket{\pm \frac{1}{2}}$ ($\ket{\pm \frac{3}{2}}$) state. Therefore, not only would the colour centre become much brighter, but also strong coupling between the spin and cavity mode is possible, enabling the spin-cavity system to serve as a deterministic spin-photon interface.
	\begin{table}[htb]
		\centering
		\begin{tabular}{c|c}
			\hline
			Parameter & Value
			\\ \hline
			Cooperativity $\left(\ket{\pm \frac{1}{2}}\right)$ & 11.2
			\\ \hline
			Cooperativity ($\left(\ket{\pm \frac{3}{2}}\right)$ & 21.0
			\\ \hline
			Cyclicity $\left(\ket{\pm \frac{1}{2}}\right)$ & 71.9
			\\ \hline
			Cyclicity $\left(\ket{\pm \frac{3}{2}}\right)$ & 103.6
			\\ \hline
			Internal Quantum Efficiency $\eta_{QE}$ $\left(\ket{\pm \frac{1}{2}}\right)$ & 0.946
			\\ \hline
			Internal Quantum Efficiency $\eta_{QE}$ $\left(\ket{\pm \frac{3}{2}}\right)$ & 0.983
			\\ \hline
			Debye Waller Factor & $0.973$
			\\ \hline
			Cavity Decay Rate $\kappa$ & $\SI{12.2}{\GHz}$
			\\ \hline
			Coupling Rate g & $\SI{5.65}{\GHz}$
			\\ \hline
		\end{tabular}
		\caption{Computed cavity enhanced spin parameters for \ce{V_2} centre.}
		\label{tab:cavity_enhancement}
	\end{table}

	We further consider the effect of cavity on spin control and readout, a requisite for realizing a quantum register. Chiefly, only readout and initialization are affected by the presence of a cavity, due to (common) utilization of resonant optical transition. In reference to figure \ref{fig:v2_energylevel}, a qubit is realized using either $\ket{+ \frac{1}{2}}$, $\ket{+ \frac{3}{2}}$ or $\ket{- \frac{1}{2}}$, $\ket{- \frac{3}{2}}$ ground states. For the addressed state (commonly $\ket{\pm \frac{3}{2}}$ for higher cyclicity), repeated optical cycling results in emission of photons (signal) from radiative decay before eventual relaxation to the opposite states through the probabilistic non-radiative Inter-System Crossing (ISC) process, while the unaddressed state remains dark in the ground state, achieving both readout and initialization. In consideration of the mechanism, primarily two cavity effects are relevant, the cyclicity enhancement that yields more signal photons for readout, and the undesirable linewidth broadening that reduces spin selectivity of excitation and leads to reduction in both signal contrast and spin polarization.  

	Both effects influence spin readout and initialization in a non-trivial manner, furthermore the impact of broadened linewidth does not scale monotonically with cavity Q factor. As illustrated in figure \ref{fig:optical_broadening}, the overlap between the two transitions increases with Q factor and peaks at $Q \sim 10^{5}$, with corresponding reduction and eventual recovery of optical spin polarization, depicted in \ref{fig:optical_initialization_fidelity}. Therefore, to understand the combined effect, we perform a numerical calculation of the readout and initialization fidelity by single-shot readout \cite{10.1038/s41467-020-20755-3} and measurement based initialization \cite{10.1088/2058-9565/ad133f}. As shown in figure \ref{fig:singleshot_fidelity}, both fidelities increase with Q factor up till around $10^{4}$, at which point the fidelities either stay roughly constant or drop. We attribute it to the reduction in signal contrast negating the improvement in cyclicity. \emph{Heuristically}, although the signal contrast reduction brings the two distributions closer together, the relative spread of each distribution reduces with increasing cyclicity, consequently keeping the overlap of the two distributions roughly similar. Beyond $10^{5}$, the increase resumes, reaching close to unity at $Q \sim 10^{6}$. Even with the achieved Q factor of $Q = 13 \times 10^{3}$, readout fidelity of $94.1 \%$ and initialization fidelities of $98.0 \%$ and $98.5 \%$ for $\ket{\pm \frac{3}{2}}$ and $\ket{\pm \frac{1}{2}}$ could be achieved, sufficient for application in quantum networks and improved spin readout.

	\section{Conclusion}
	In this work, we proposed and realized a tunable hybrid material cavity design, made in part from lithium niobate, and successfully demonstrated electrical tuning of the fabricated devices. We further evaluated and showed the prospect of utilizing the cavity for quantum networking and spin readout applications. The proposed technique presents a scalable approach towards realization of quantum networks based on solid state qubits. Future work includes the incorporation of spins into the cavity, experimentation with other spin hosting materials, like diamond, as well as exploration of other cavity designs.

\section{Acknowledgement}
	S.L.N thanks Ruoming Peng for fruitful discussions and support, Mo Li group at the University of Washington for computational support with simulation, Jonathan Körber, Di Liu, Fiametta Sardi for fruitful discussions, Yong Lu for early guidance. We acknowledge financial support from European Union’s Horizon Europe under project SPINUS (Grant No. 101135699), German Federal Ministry of Education and Research under project QR.N (Grant No. 16KIS2207), German Federal Ministry of Research, Technology and Space under project QSi2V (Grant No. 13N16756) and project LichtBriQ (Grant No. 16KISQ031) and Center for Integrated Quantum Science and Technology.

\bibliographystyle{unsrtnat}
\bibliography{Citation}
\end{document}